\documentclass{osa-article}

\journal{osac}

\articletype{Research Article}

\usepackage{lineno}
\usepackage{siunitx}

\begin{document}

\title{High-Performance Photon Number Resolving Detectors for 850-950~nm wavelengths}

\author{J. W. N. Los\authormark{1,*}, Mariia Sidorova\authormark{2,6}, B. L. Rodriguez\authormark{3}, Patrick Qualm\authormark{3},  J. Chang\authormark{4}, S. Steinhauer \authormark{5}, V. Zwiller\authormark{5}  and I. Esmaeil Zadeh\authormark{3}}

\address{\authormark{1}Single Quantum B.V., 2628 CH Delft, The Netherlands
\\
\authormark{2}School of Physical \& Mathematical Science, Nanyang Technological University, 21 Nanyang Link, SPMS-PAP-02-12c, 637371, Singapore
\\
\authormark{6}German Aerospace Center (DLR)$,$ Institute of Optical Sensor Systems$,$ Rutherfordstr. 2$,$ 12489 Berlin$,$ Germany
\\
\authormark{3}Department of Imaging Physics,  Delft University of Technology, 2628CN Delft, The Netherlands.
\\
\authormark{4}Kavli Institute of Nanoscience, Department of Quantum Nanoscience,
Delft University of Technology, 2628CJ Delft, The Netherlands
\\
\authormark{5}Department of Applied Physics, Royal Institute of Technology (KTH), SE-106 01 Stockholm, Sweden}

\email{\authormark{*}niels@singlequantum.com} 



\begin{abstract}
Since their first demonstration in 2001 \cite{goltzman2001}, superconducting-nanowire single-photon detectors (SNSPDs) have witnessed two decades of great developments. SNSPDs are the detector of choice in most modern quantum optics experiments and are slowly finding their way into other photon-starved fields of optics. Until now, however, in nearly all experiments SNSPDs were used as 'binary' detectors, meaning they can only distinguish between 0 and \(>=\)1 photons and photon number information is lost. Recent research works have demonstrated proof-of-principle photon number resolving (PNR) SNSPDs counting 2-5 photons. The photon-number-resolving capability is highly demanded in various quantum-optics experiments, including Hong–Ou–Mandel interference, photonic quantum computing, quantum communication, and non-Gaussian quantum state preparation. In particular, PNR detectors at the wavelength range of 850-950~nm are of great interest due to the availability of high-quality semiconductor quantum dots (QDs) \cite{Heindel_QD_review:2023} and high-performance Cesium-based quantum memories\cite{Ma_QMemory:2017}. In this paper, we demonstrate NbTiN-based SNSPDs with > 94\% system detection efficiency, sub-11~ps timing jitter for one photon, and sub-7~ps for 2-photon. More importantly, our detectors resolve up to 7 photons using conventional cryogenic electric readout circuitry. Through theoretical analysis, we show that the current PNR performance of our detectors can still be further improved by improving the signal-to-noise ratio and bandwidth of our readout circuitry. Our results are promising for the future of optical quantum computing and quantum communication.
\end{abstract}

\section{Introduction}
Photons, owing to their unique characteristics, serve as promising candidates for various quantum experiments and applications. Modern lasers are capable of producing monochromatic, coherent and highly directional light enabling the precise manipulation of photons and matter. Together with the state-of-the-art single-photon emitters \cite{Aharonovich_single_photon_emitters:2016,Zhai_2photon_goodQD:2022}, and high-performance single-photon detectors, they provide a promising avenue for the future implementation of quantum computers and networks.\cite{leonhardt2010essential}. {With classical optics being a well-established theory, the comprehensive understanding of its classical aspects allows us to direct our attention toward exploring the non-classical quantum effects \cite{fabre2020modes}}. This understanding has prompted numerous fundamental tests of quantum mechanics within the realm of quantum optics as well as applications, including reconstructing arbitrary light source statistics \cite{classen2016superresolving}, quantum communication and cryptography \cite{lutkenhaus2002quantum}, non-gaussian quantum state preparation \cite{menzies2009gaussian}, quantum-enhanced imaging \cite{meda2017photon}, LiDAR \cite{cohen2019thresholded} and photonic quantum computing \cite{PhysRevLett.131.150601}. Among these cutting-edge experiments, photon-number resolution (PNR), which refers to the capability of a detector to resolve the number of incident photons that are closely spaced in time, plays a crucial role. 

Among all technologies for single-photons detection, superconducting nanowire single-photon detectors (SNSPDs) \cite{esmaeil2021superconducting,chang2023nanowire} have demonstrated superior performance in terms of system detection efficiency (>98\%) \cite{chang2021detecting,reddy2020superconducting,hu2020detecting}, low timing jitter (<10~ps) \cite{Korzh_3ps_jitter:2020,esmaeil2020efficient}, low dark count rates (10$^{-4}$~Hz) \cite{zhang2011ultra} and high count rates (>100~Mcps) from the visible to the mid-infrared range \cite{chang2019multimode,cirillo2020superconducting,chang2022efficient,taylor2023low}. Recently, the operating temperature of SNSPDs has started to gradually increase \cite{chang2023superconducting,charaev2023single,merino2023two}, making SNSPDs even more attractive for applications. Nevertheless, SNSPDS, so far, were mostly used as binary detectors, distinguishing photon numbers zero and higher than zero. In the past few years, a few works have demonstrated photon number resolving (PNR) capabilities \cite{cahall2017multi, zhu2020resolving, stasi2023fast}. The combination of high-resolution PNR with high detection efficiency and time resolution would enable a plethora of applications. The wavelength range of 850-950\,~nm is of particular interest as it includes some of the most promising quantum emitters as well as high-performance Cesium-based quantum memories \cite{reimer2012bright, he2023single}. 

\section{Basic concept and measurement schemes}

Upon photon absorption in a superconducting nanowire, the photon energy is transferred to electrons and phonons. Since this energy is orders of magnitude greater than the binding energy of the Cooper pairs (given by the superconducting gap), a small non-superconducting region known as a 'hot spot' is created. Whether by diffusion of quasi-particles or nucleation and dynamics of vortices, a normal domain across the wire is formed, and the wire undergoes a transition into the resistive state. This blocks the superconducting current, resulting in a voltage spike that is then amplified and registered by the readout circuit. Subsequently, a relaxation process restores the superconductivity in the device. 

\begin{figure}[ht!]
\centering\includegraphics[width=14cm]{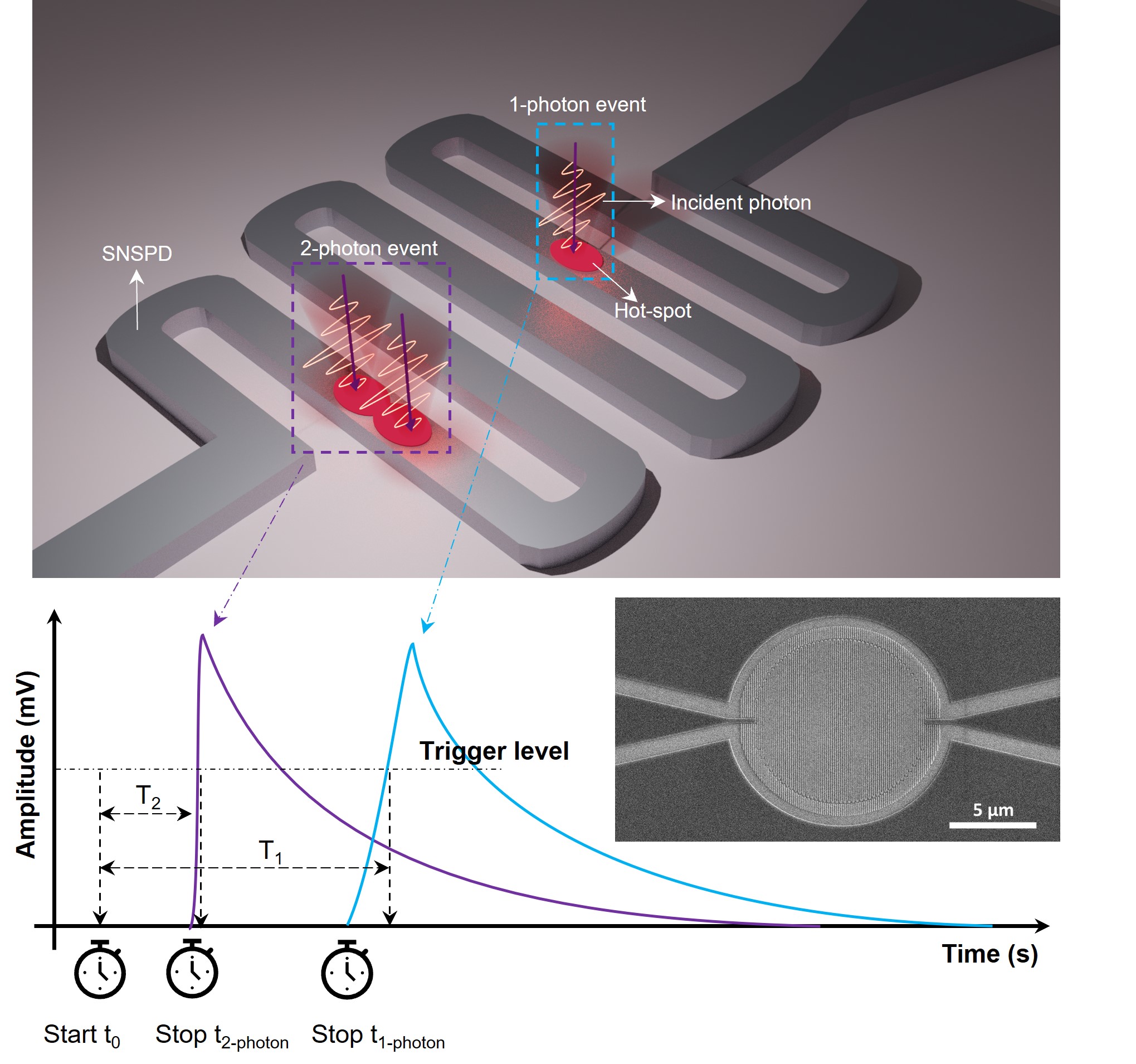}
\caption{An illustration of a SNSPD impinged by multiple photons. The number of photons that are absorbed by the detector modifies the rising edge of the detection pulse and hence can be used to resolve the photon number. Inset: Scanning Electron Microscope image of a similar SNSPD similar to the ones used in this study. }
\label{fig:illustration}
\end{figure}

When multiple photons arrive, several hot spots are created along the nanowire. However, usual readout schemes for SNSPDs are not able to show a difference in the output signal. By using a low noise amplifier, a signal proportional to the number of photons can be detected in the oscilloscope. Using an appropriate signal processing protocol, photon-number resolution can be achieved.

\section{Intrinsic PNR capabilities of SNSPD}
To explore the intrinsic ability of SNSPD to resolve the photon number encoded in the rising edge of its pulses, we first consider aspects that are crucial for accurate photon-number resolution and then outline practical guidelines for improving SNSPDs' PNR efficiency.

\subsection{Counting statistics}
To correctly reconstruct the statistics of incident light using SNSPD one has to account for: (i) the probability $P_\eta^N(n|q)$ that $n\leq q$ photons are detected from $q$ incident photons, (ii) the probability $P(n|n+1)$ to discriminate between the arrival times of SNSPD's pulses initiated by $n$ and $n+1$ photons. The latter is particularly important to avoid underestimating events with higher photon numbers. 

\begin{figure}[h!]
\centering\includegraphics[scale=0.9]{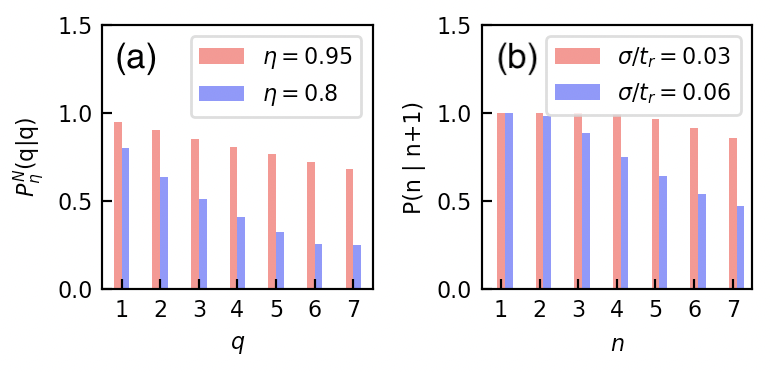}
\caption{(a) The probability that all incident photons are detected. (b) The probability of discriminating between the arrival times of detector pulses initiated by $n$ and $n+1$ photons for two $\sigma/t_R$ ratios is indicated in the legend.}
\label{fig:probabilities}
\end{figure}

A common approach is to treat a uniformly illuminated device as a spatially multiplexed N-element array ($N \approx 10^3$) of identical detectors with uniform detection efficiency ($\eta$). Here, the size of each independent element is defined by the length of a normal domain ($\approx 1\mu$m). For $q \ll N$, the probability that more than one photon will hit the same element is very small ($<1\%$ for $q=5$ and $<5\%$ for $q=10$), and the probability in  (i) is strongly influenced by the detection efficiency $\eta$. This probability has been derived in previous works (e.g., in \cite{fitch2003photon}): $P_\eta^N (n|q)=\dfrac{N!}{n! (N-n)!}\sum_{j=0}^n (-1)^j \dfrac{n!}{j!(n-j)!}\left[(1-\eta)+(n-j)\dfrac{\eta}{N}\right]^q$. It incorporates non-ideal efficiency $\eta<1$ and accounts for the likelihood of multiple photons hitting the same element (fig.~\ref{fig:probabilities}a). Given the normally distributed arrival times of the detector's voltage pulses, the probability in (ii) accounts for the overlapping coefficient between the normal distributions of their respective pulse, given by $P(n|n+1) = \frac{1}{2} \left[erf\left(\dfrac{c-\mu_{n+1}}{\sigma_{n+1}\sqrt{2}}\right) - erf\left(\dfrac{c-\mu_n}{\sigma_n\sqrt{2}}\right)\right]$ (fig.~\ref{fig:probabilities}b). Here, $c$ is the distributions’ intersection point, $erf(x)$ is the error function. The mean $\mu$ and the standard deviation $\sigma$ are linked to the pulse rising time $t_R$ and the jitter $\sigma$. According to the electrothermal model (similar to \cite{yang2007modeling} with non-linear thermal equations as in \cite{sidorova2022phonon}), $t_R\propto n^{-0.3}$ and $\sigma \propto t_R$. Reducing the $\sigma/t_R$ ratio allows for near-ideal photon-number resolution (see fig.~\ref{fig:probabilities}b; the limited amplifier bandwidth will result in a sharp cut-off, not shown).

\subsection{Strategies for enhancing PNR efficiency}
Analysis of the counting statistics has revealed that, to achieve a near-ideal photon-number resolution, two key parameters to be optimized are: detection efficiency and the $\sigma/t_R$ ratio. Considering the noise-dominated experimental jitter, the implementation of low-noise, broad-bandwidth cryogenic electronics becomes crucial. Analysis of the electron-thermal model indicates that the SNSPD's PNR efficiency is tied to the lifetime of the normal domain. This correlation reveals specific material and geometric parameters, which control domain lifetimes and can optimally enhance the intrinsic PNR efficiency. Particularly beneficial are: (a) increasing the nanowire length, (b) reducing its sheet resistance, (c) reducing the nanowire width, and (d) increasing the acoustic mismatch between the film and substrate. In addition, materials with a higher critical temperature can offer lower intrinsic timing jitter, thus improving PNR capabilities.

In scenarios, when photons do not hit the device simultaneously but arrive with a delay, their number can still be resolved. The maximum delay between photons that still allows the resolution of their number is determined by a combination of the domain lifetime and the current through the device. When the domain lifetime reaches about one-third, the current decreases by half, resulting in lower detection efficiency (fig.\ref{fig:pnr_events}b). Strategies focusing on increasing the domain lifetime (a,b,d) are also beneficial in this case. It is worth mentioning that electro-thermal models relying on a propagating domain wall (e.g., in \cite{kerman2009electrothermal}) can also be applicable for the case of delayed domains (e.g., one has to double the domain's wall velocity at the moment when the second photon is absorbed).

\section{Experimental methods}

Superconducting single-photon detectors were made from NbTiN thin films deposited by DC Magnetron Sputtering. The thickness of the NbTiN film is 10~nm, and it is patterned into a  70~nm wide nanowire with a period of 140~nm (corresponding to a fill factor of 50\%) covering a circular area with a radius of 6~$\mu$m. See inset of figure \ref{fig:illustration} for an SEM image of an SNSPD. The devices are fabricated on top of a DBR stack to maximize the absorption around a wavelength of 940~nm. The distance between the detector and the tip of the fiber is controlled using non-deformable metal spacers with their thickness determined by FDTD simulations (Lumerical). 
The detectors are tested in a Gifford-McMahon Cryocooler with an operating temperature of 2.5~K \cite{chang2022efficient}, well below the critical temperature of the superconducting films, and are optically addressed using a polarization-maintaining fiber.  

\begin{figure}[ht!]
\centering\includegraphics[scale=0.4]{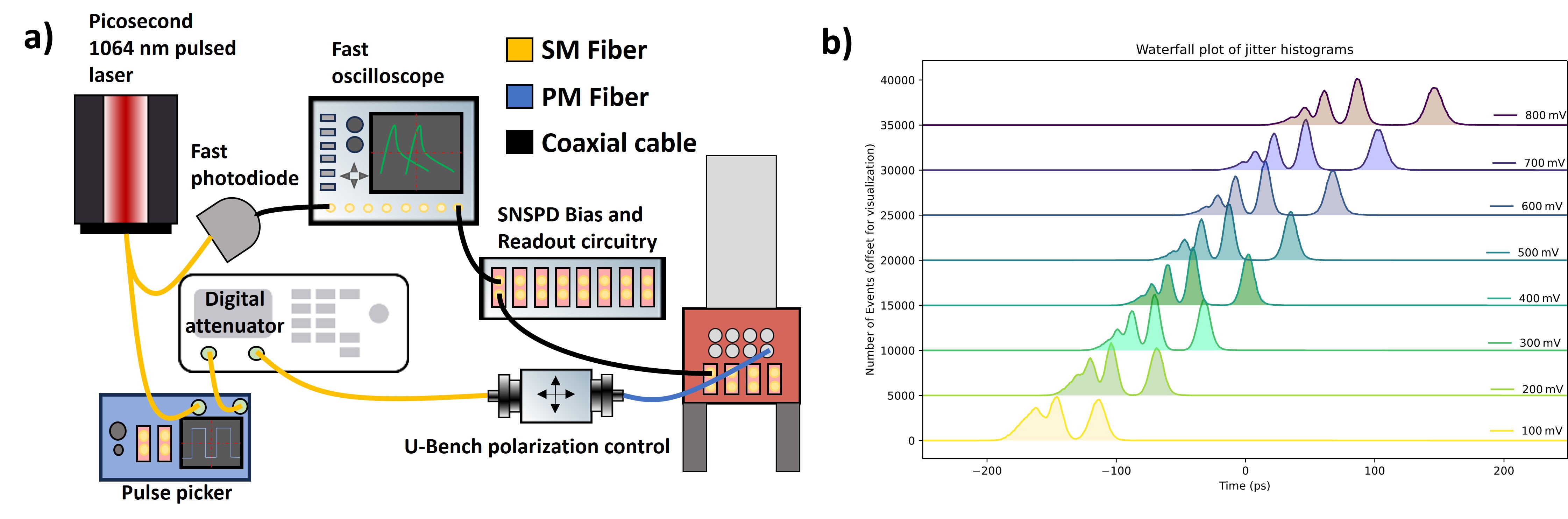}
\caption{\label{fig:pnr_setup}(a) Experimental setup for testing PNR capabilities of studied SNSPDs. (b) Waterfall plot of jitter measurements for trigger level sweep to determine the optimal setting for the PNR measurement, achieved at 700~mV.}
\end{figure}

A schematic of the experimental setup used during these experiments can be seen in fig.\ref{fig:pnr_setup}a. As a continuous wave light source, we used a tunable laser (SpectraPhysics Millenia eV Model 3910) and for the pulsed measurements, a 1064 nm picosecond laser (Ekspla with a pulse length of 2.3 ps) with a pulse picker allowing pulse repetition rates of 20~kHz - 40~MHz. The pulsed measurements are performed at a repetition rate of 1012~kHz to ensure the events are well separated in time and therefore do not influence each other.  

The detectors are DC-biased at about 0.97 of the switching current (I$_{sw}$= 37.5 $\mu$A) while the one with the highest efficiency is biased at around 17.5~$\mu$A (I$_{sw}$= 18 $\mu$A)). At this bias current, the dark count rate of the detectors was about 40~cps. The device is connected to a low-noise cryogenic amplifier (about 2~GHz bandwidth) using cryogenic coaxial cables (3~dB/m loss at 1~GHz). An oscilloscope (Lecroy waverunner 40~GS/s and 4~GHz bandwidth) is used to perform the timing jitter measurements, where we also measure the distribution between the detector pulse and the output signal of a fast reference photodiode coming from the pulsed laser. In this setup, the jitter and photon-number-resolving behavior of the detector are investigated. For the extraction of the graphs and representation, the signal waveform of the oscilloscope is processed using a customized MATLAB script. For the measurements the pulse count rate is fixed, and consequently a timing jitter measurement is performed until 200k events are collected. This is repeated for a range of skew levels to determine at what trigger level the optimal photon-number-resolving results are achieved. Fig.\ref{fig:pnr_setup}b shows a plot of the photon events occurring at different times obtained by varying the trigger level configurations, from 100~mV to 800~mV in steps of 100~mV. The later photon arrival time for higher trigger levels in the waterfall plot is a result of the pulse rise time. 

\section{Photon number resolution up to 7 photons}

\begin{figure}[ht!]
\centering\includegraphics[scale=0.4]{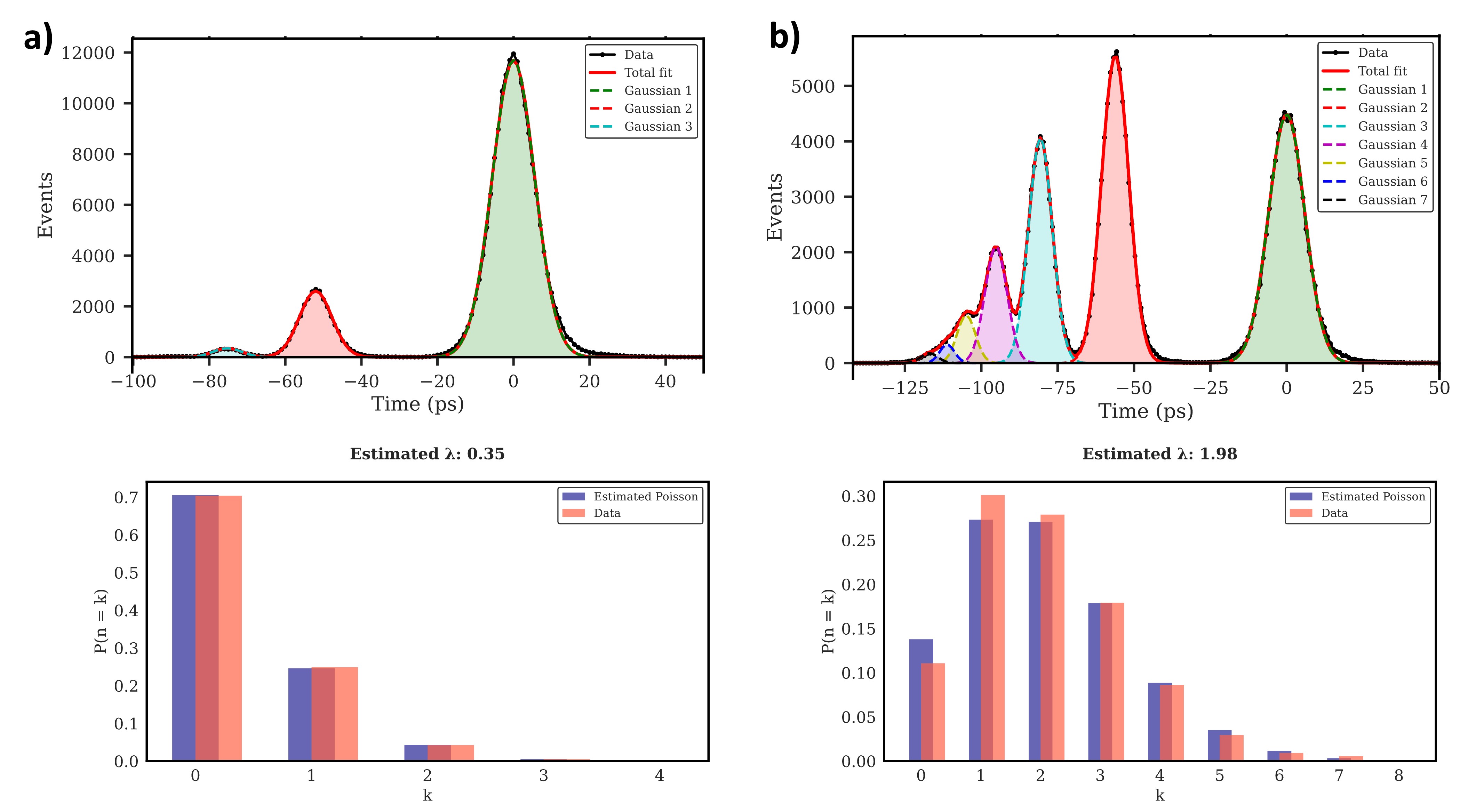}
\caption{\label{fig:pnr_events}PNR experiments at trigger level of 700~mV and average pulse count rate of (a) 350~kcps and (b) 900~kcps and their corresponding reconstructed photon statistics.For both measurements the laser repetition rate is fixed at 1012~kHz. The measured photon number zero is, contrary to expectations, slightly lower than the expected value which we ascribe to the low extinction ratio of our pulse picker, effectively causing an increase in measured photon number one.}
\end{figure}

As discussed before, we used a picosecond pulsed laser at 1064\,nm to characterize the PNR capabilities of our detectors. Since our detectors are optimized for 850-950\,nm (DBR cavity optimized for that wavelength range), the absorption and hence the efficiency is significantly lower at 1064\,nm ($\sim$40\%). Therefore, we verified the photon number resolving capabilities of our detectors using a method which does not require taking into account detector efficiency (a more practical approach) as follows: 1) we use a pulsed laser and create arrival time histograms similar to Fig.\ref{fig:pnr_setup}b. 2) We ascribe each peak to a photon number, e.g. 1, 2 etc. (an ansatz). 3) We calculate the total number of detected photons (equivalent with the number of absorbed photons if internal detection efficiency is unity) by integrating over the curve of each peak and multiplying it by the assumed photon number, the outcome is then the total number of detected photons. 4) Given the known number of detected photons and the repetition rate of the laser, we can calculate the average detected photon number per pulse and hence make a prediction for expected photon number distribution (based on Poisson distribution). 5) Finally, we contrast the predicted versus measured distributions. Examples are shown in  Fig.\ref{fig:pnr_events}. While, as demonstrated in Fig.\ref{fig:pnr_events}, the predicted versus the measured photon numbers are in good agreement, it is clear that the distribution for photon number zero is overestimated while the peaks for photon numbers of 1 and 2 are slightly underestimated in contradiction with expected behaviors shown in Fig.\ref{fig:probabilities}. These discrepancies arise from the relatively poor extinction ratio of our pulse picker (20dB, see supporting information).

The detector with the best photon-number resolving capabilities was analyzed at different photon fluxes. The optimal trigger level in which 7 photon events can be distinguished is found at 700~mV. In Fig.\ref{fig:pnr_events}a and Fig.\ref{fig:pnr_events}b multiple photon events can be resolved, from 3 to 7 photons corresponding to the used photon fluxes of 350~kcps and 900~kcps respectively. Using known photon statistics, it is possible to reconstruct the expected behaviour and compare it with the experimentally obtained data. Reconstructing the photon statistics using the Poissonian distribution shows that the data matches the theoretical estimation with high overlap. At small mean photon numbers ($\lambda=0.35$ in Fig.\ref{fig:pnr_events}a), the estimated Poisson probabilities is nearly identical to those obtained in the experiments. However, at higher mean photon numbers ($\lambda=1.98$ in Fig.\ref{fig:pnr_events}b), there are discrepancies, especially for photon numbers $k=0-2$. We attribute this to the finite suppression of the unwanted pulses by the laser pulse picker in our setup, an issue that deteriorates as the the mean photon number increases.

\section{High system detection efficiency and photon number resolution}

\begin{figure}[h!]
\centering\includegraphics[width=13cm]{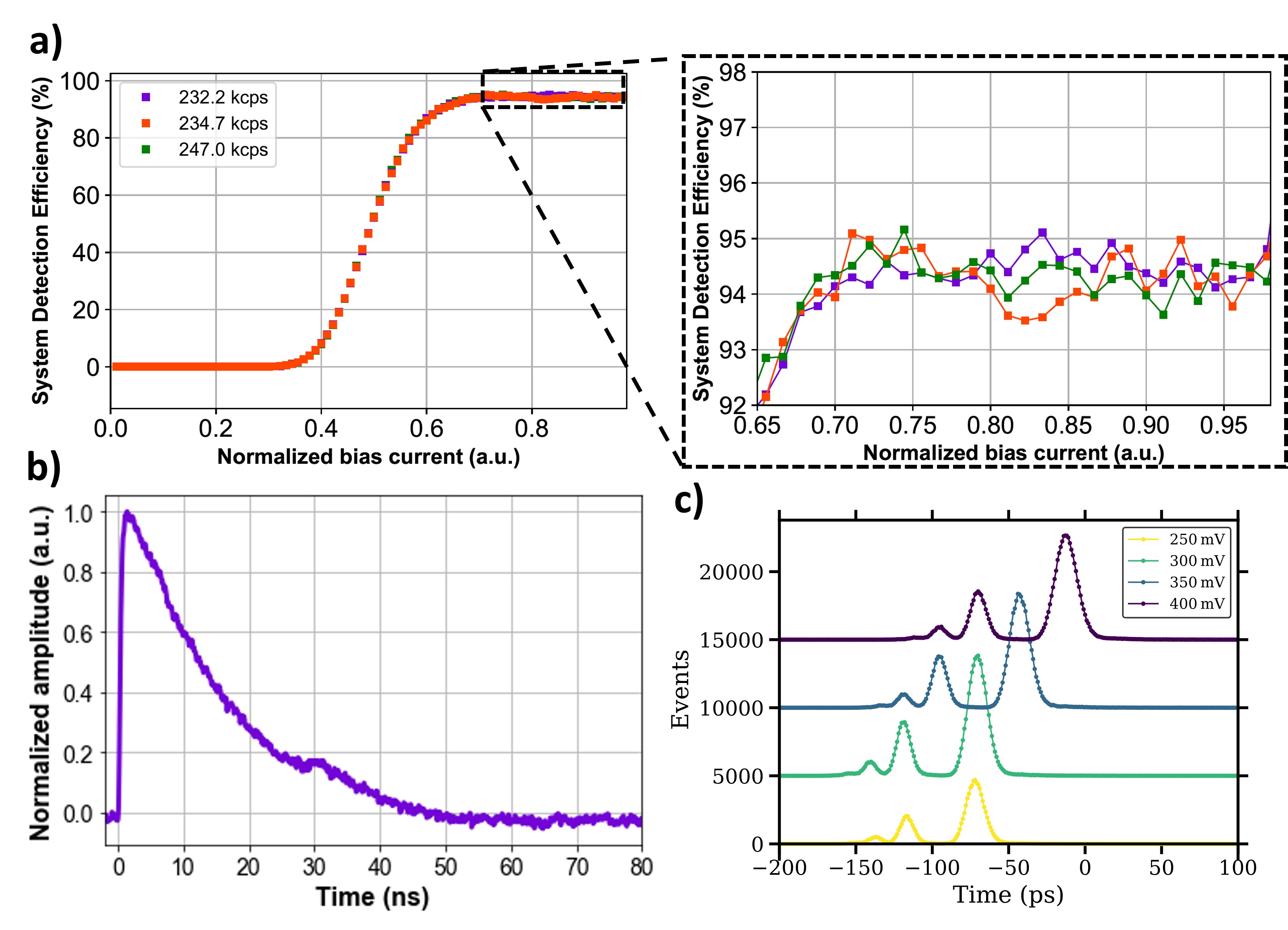}
\caption{\label{fig:efficiency_pnr}a) System detection efficiency curves of the detector under study with three similar count rates. We measured an average efficiency of 94.5$\%$ at the wavelength of 940 nm taken from three consecutive measurements. b) Typical voltage pulse from the studied detector and c) Photon Number Resolving capabilities of the detector with different voltage skew levels demonstrating different photon events}
\end{figure}

Another detector with high efficiency at 940 nm was analyzed to show that photon number resolution behaviour can also be achieved and the results from this study are shown in fig.\ref{fig:efficiency_pnr}. To measure the efficiency of the detector, the photon flux going to the detector was determined with a reference arm, and the relative optical power between these fibers was measured by using two Newport power-meters, as explained in our previous work \cite{chang2021detecting}. The system detection efficiency of the detector was measured at \SI{940}{\nano\meter} with an optical power between 1.88-2.00~nW measured using a NIST traceable power-meter (818-SL-L Newport, uncertainty 1.1$\%$). Consequently, the power was attenuated by around \SI{45}{\deci\bel} with free-space OD plates. The attenuation was determined by measuring the power from a reference arm using another Newport power-meter (818-IR Newport, linearity 0.5$\%$). The averaged SDE over three measurements, shown in fig.\ref{fig:efficiency_pnr}a, was 94.5$\%$. In these measurements we subtracted 3.4$\%$ that is the fiber-air reflection and causes underestimation of the photon flux, calculated using the transfer matrix method. The measurement uncertainty can be inferred from the power-meter uncertainty (1.1$\%$) and its linearity uncertainty (0.5$\%$), giving an error of 1.3$\%$ in the system detection efficiency. The photon count rate behavior as a function of the bias current is shown in fig.\ref{fig:efficiency_pnr}a. For this detector, the time jitter at \SI{1064}{\nano\meter} was 18.4~ps.
 
A typical graph of an histogram of photon events as a function of time is shown in figure \ref{fig:efficiency_pnr}c. 
The visibility of the photon peaks can be optimized by adjusting the skew level to an optimal value. By fitting the time response of the consecutive histograms to Gaussian-shaped curves, four peaks can be resolved when the skew level is 200~mV.

\section{Conclusions}

Superconducting nanowire single-photon detectors with high efficiencies (around 94.5$\%$ at 940 nm) and high timing resolution (18.4 ps at 1064 nm) have demonstrated photon-number-resolving behavior. Most important, for a detector of similar characteristics we demonstrate photon-number-resolution up to 7 photons corroborated by applying photon statistics. In addition, we propose strategies to further expand PNR capabilities towards higher photon numbers by optimizing the detector material properties and geometric parameters. In conclusion, the demonstration opens the path to interesting experiments with these detectors in photonic quantum computing, quantum communication, and in combination with semiconductor quantum dots and Cesium-based quantum memories.

\section{Acknowledgments}
J. W. N. L. and I. E. Z. acknowledge funding from the European Union’s Horizon Europe research and innovation programme under grant agreement No 101098717 (RESPITE project) and number 101099291 (fastMOT project).

M.S. acknowledges funding support from the National Research Foundation, Singapore and A*STAR under the Quantum Engineering Pro-gramme (QEP-P1).

\section{Data Availability Statement}
The data that support the findings of this study are available on request from the authors.

\bibliography{sample}

\end{document}